\documentclass{article}

\usepackage[preprint,nonatbib]{neurips_2026}
\usepackage{amsmath,amssymb,amsfonts,bm}
\usepackage{booktabs}
\usepackage{multirow}
\usepackage{graphicx}
\usepackage{url}
\usepackage{hyperref}
\usepackage{xcolor}
\newcommand{\metriccell}[1]{\scriptsize #1}
\usepackage{array}

\newcommand{\cope}{\textbf{COPE}}

\title{Opening-Price Shortcut in Stock Prediction: A Conditional Opening Prior-and-Evidence Framework}

\author{%
Zhengyang Fang \\
Beijing University of Posts and Telecommunications \\
\And
Zhongliang Yang \\
Beijing University of Posts and Telecommunications \\
\And
Linna Zhou \\
Beijing University of Posts and Telecommunications
}

\begin{document}

\maketitle

\begin{abstract}
Stock price prediction has long been a central problem in quantitative finance. While recent methods increasingly leverage external modalities such as news to enhance predictive performance, the microstructure within price sequences itself contains crucial information. The target-day opening price represents the first concentrated realization of overnight information in the market and holds significant value for intraday evolution and the eventual closing direction. However, once opening information is introduced, the opening and closing directions often exhibit high correlation, making the model prone to degenerating into an opening-price shortcut that simply extrapolates the closing direction from the opening direction. To address this, we propose \cope{} (\textbf{C}onditional \textbf{O}pening \textbf{P}rior and \textbf{E}vidence), a shortcut-aware prediction framework that relies solely on the price modality. \cope{} models the target-day opening state as an observed conditioning variable and reparameterizes the closing-direction prediction into a continuation/reversal discrimination conditioned on the opening direction. Furthermore, we decompose the input evidence into opening condition, individual historical state, and contextual support, and perform residual modeling of historical and contextual evidence under this condition to distill information truly effective for the final closing judgment. Systematic experiments on four real-market datasets demonstrate that \cope{} significantly outperforms existing methods. Ablation studies and shortcut-sensitive analyses further validate the effectiveness of explicitly modeling opening information, while opening-time backtests reveal that predictive accuracy and trading returns align.
\end{abstract}

\section{Introduction}
\label{sec:intro}

Stock direction prediction is a core task in quantitative finance and financial technology: given historical market observations, the model must determine the price direction on the target trading day~\cite{nelson2017stock,qin2017dual,ijcai2019p810}. Existing research has extended from single-stock historical price sequences~\cite{yoo2021accurate,fan2024stockmixer,jiang2021applications} to cross-stock relations, news, social media, earnings reports, and large language model-assisted external information~\cite{0News,2001News,2014News,schumaker2009textual,si2013exploiting,xing2018natural,xu2018stock,hu2018listening,luo2023causality,li2024causalstock,li2023pen,wang2024llmfactor}, yet the microstructure within price sequences still warrants further modeling.

The target-day opening price is the first concentrated price realization of overnight information within the trading day~\cite{aboody2018overnight,lou2019tug,ajello2023more}; it is already observable at the market open and is naturally correlated with the eventual adjusted-close direction. Directly concatenating the opening direction or opening magnitude into the model input may yield apparent accuracy gains, but such improvements need not stem from better historical-state understanding---the model may instead degenerate into a low-complexity discriminator that extrapolates the

\begin{figure}[t]
\noindent
\raisebox{0pt}[0pt][0pt]{\makebox[\linewidth][r]{%
\begin{minipage}[t]{0.5\linewidth}
\vspace{0pt}
\centering
\includegraphics[width=\linewidth]{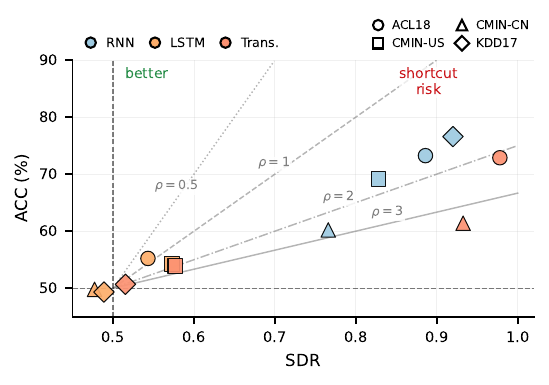}
\caption{Opening-price shortcut pre-experiment. The plot shows ACC, SDR, and $\rho$ contour lines for direct-opening-input models on four datasets.}
\label{fig:intro_shortcut}
\end{minipage}}}
\parshape=24
0pt 0.43\linewidth
0pt 0.43\linewidth
0pt 0.43\linewidth
0pt 0.43\linewidth
0pt 0.43\linewidth
0pt 0.43\linewidth
0pt 0.43\linewidth
0pt 0.43\linewidth
0pt 0.43\linewidth
0pt 0.43\linewidth
0pt 0.43\linewidth
0pt 0.43\linewidth
0pt 0.43\linewidth
0pt 0.43\linewidth
0pt 0.43\linewidth
0pt 0.43\linewidth
0pt 0.43\linewidth
0pt 0.43\linewidth
0pt 0.43\linewidth
0pt 0.43\linewidth
0pt 0.43\linewidth
0pt 0.43\linewidth
0pt 0.43\linewidth
0pt 0.43\linewidth
\noindent closing direction from the opening direction. We term this phenomenon the \emph{opening-price shortcut}. In Figure~\ref{fig:intro_shortcut} we construct a simple yet effective pre-experiment, where higher ACC is better, and lower SDR and $\rho$ are better (detailed definitions are provided in the Experiments section). The results show that conventional models, upon receiving additional opening-price features, struggle to balance ACC against SDR and $\rho$ across multiple datasets, confirming that naively introducing opening information without careful analysis indeed induces the opening-price shortcut. Meanwhile,\par
\vspace{0.45em}
\end{figure}

underpowered models such as LSTM fail to exploit the shortcut and remain at chance-level accuracy. The opening-price shortcut problem thus demands formalization and targeted solutions.

Therefore, this paper addresses how to leverage the opening state while avoiding mechanically copying the opening direction as the final prediction. We formalize this risk as a measurable problem: inspired by shortcut learning~\cite{steinmann2024navigating,zheng2025shortcutprobe} and spurious correlation~\cite{ye2026the,zheng2024learning}, we define task-specific metrics for the opening-price shortcut. The Shortcut Direction Rate (SDR) measures the proportion of predictions that agree with the opening direction, and the shortcut dependence ratio $\rho$ quantifies the strength of this agreement relative to above-chance accuracy gains, thereby distinguishing genuine predictive improvement from opening-direction copying.

Based on the above observations, we propose \cope{}, a conditional opening prior-and-evidence framework that conditions on the target-day opening state. \cope{} does not change the ultimate task---it still predicts the direction of the target-day adjusted close relative to the previous day's adjusted close---but it changes the pathway by which the model arrives at this prediction. Specifically, it treats the opening direction as an observed conditioning variable and reparameterizes the original direction label into an equivalent continuation/reversal discrimination, so that the model judges whether the opening move will be corrected before the close. \cope{} extracts residual evidence relative to the opening-conditioned prior through opening-conditioned prior, historical state, market contextual summary, reliability gating, and K-conditioned residual attention, and recovers the final closing direction via $Y=K\oplus R$.

The remainder of this paper is organized as follows. Section~\ref{sec:related} reviews related work and positions our contribution. Section~\ref{sec:method} formalizes the opening-price shortcut and presents \cope{}\'s conditional reparameterization, opening-conditioned prior, reliability gating, K-conditioned residual attention, and training objective. Section~\ref{sec:experiments} reports main comparisons, ablations, shortcut-sensitivity analyses, and backtests. Section~\ref{sec:conclusion} summarizes findings and discusses future directions.

Our contributions are summarized as follows:
\begin{itemize}
    \item We are the first to identify the opening-price shortcut as a measurable, reproducible research problem under the target-day opening state, and we propose diagnostic metrics compatible with the closing-direction prediction task.
    \item We propose \cope{}, which equivalently reparameterizes the adjusted-close direction prediction into a continuation/reversal discrimination conditioned on the opening state. By first interpreting the opening state as a conditional prior, \cope{} shifts the prediction focus from copying the opening direction to judging whether the opening move will be corrected by current trading evidence.
    \item We design systematic experiments on four real-market datasets to evaluate predictive performance, shortcut suppression capability, and trading application value. Experimental results demonstrate significant advantages of our approach.
\end{itemize}

\section{Related Work}
\label{sec:related}

\subsection{Unimodal Price-Sequence Prediction}
Unimodal stock prediction primarily learns temporal patterns from historical price and volume sequences. Early deep-learning approaches typically employ recurrent neural networks and their variants, leveraging gating structures or attention mechanisms to capture long- and short-term dependencies within price sequences. ALSTM~\cite{qin2017dual} introduces attention mechanisms into financial time-series modeling, and Adv-ALSTM~\cite{ijcai2019p810} further enhances representation robustness through adversarial training. More recently, researchers have explored stronger sequence or structural modeling modules, such as attention-based multivariate time-series models~\cite{shih2019temporal}, multi-frequency trading pattern modeling~\cite{zhang2017stock}, DTML~\cite{yoo2021accurate}, and StockMixer~\cite{fan2024stockmixer}. These methods confirm that predictable patterns exist within price sequences, but they generally treat open, high, low, close, and volume as homogeneous historical inputs without explicitly distinguishing or modeling how their internal microstructural features affect the eventual closing direction.

\subsection{Multimodal and Relation-Enhanced Stock Prediction}
Multimodal stock prediction seeks to complement price sequences with information from news, social media, announcements, earnings reports, or cross-stock relations. StockNet~\cite{xu2018stock} uses generative latent variables to model the influence of textual information on stock volatility; HAN~\cite{hu2018listening} extracts stock-relevant signals from news via hierarchical attention; subsequent work further incorporates social media text, corporate relationships, graph attention, or macro--industry--stock indicators~\cite{sawhney2020deep,kim2019hats,wang2021coupling,DBLP:conf/iconip/GongE21}. CMIN~\cite{luo2023causality}, CausalStock~\cite{li2024causalstock}, MagicNet~\cite{luo2025magicnet}, PEN~\cite{li2023pen}, and LLMFactor~\cite{wang2024llmfactor} further introduce causal relations, dynamic graph structures, event information, or large language model-assisted factors. Related financial text models and general pretrained language models have also been used to construct financial semantic representations~\cite{devlin2018bert,liu2019roberta,araci2019finbert,yang2023fingpt}, enhancing cross-stock and cross-modal modeling capacity. Unlike these methods, this paper returns to the price modality itself and investigates how the model should extract effective information from the opening state when the target-day opening price is observable at prediction time.

\subsection{Overnight Information and Opening-State Modeling}
The opening price reflects the first concentrated market realization of overnight information and thus has a natural connection to the eventual closing direction. Prior work has studied overnight returns, overnight news, cross-market spillovers, or target-day opening-state prediction, including financial research on overnight returns and price discovery~\cite{aboody2018overnight,lou2019tug}, opening-state modeling methods combining LSTM with relational graph convolutions~\cite{li2021modeling}, and studies exploiting global market timezone differences to capture how earlier-opening markets affect later-opening ones~\cite{gao2022forecasting}. However, these works typically treat the opening state as a valuable signal to be predicted or exploited, without analyzing the opening-price shortcut as a measurable model-degradation risk.

\section{Problem Definition}
\label{sec:problem}

Let $X^{\mathrm{his}}_{i,t}$ denote the historical-window observation for stock $i$ on trading day $t$, containing only price features available before the close of the previous day; the target-day opening state is given in raw-price terms. Denote the raw opening price, raw closing price, and adjusted closing price as $O_{i,t}$, $C_{i,t}$, and $\tilde C_{i,t}$, respectively. Each sample input is
\begin{equation}
X_{i,t}=\left(X^{\mathrm{his}}_{i,t}, O_{i,t}, K_{i,t}\right),
\end{equation}
where $K_{i,t}=\mathbf{1}(O_{i,t}>C_{i,t-1})$ indicates the target-day opening direction. If the previous day\'s closing price is missing, or the absolute change in the target-day adjusted close relative to the previous day falls below $0.003$, the stock--day sample is filtered out. The supervision target is the adjusted-close direction:
\begin{equation}
Y_{i,t}=\mathbf{1}\!\left(\tilde C_{i,t}>\tilde C_{i,t-1}\right).
\end{equation}
Thus, the task is to predict the direction of the adjusted closing price, conditioned on the historical window and the observable target-day opening state.

\section{Method}
\label{sec:method}

\begin{figure}[t]
  \centering
  \includegraphics[width=\linewidth]{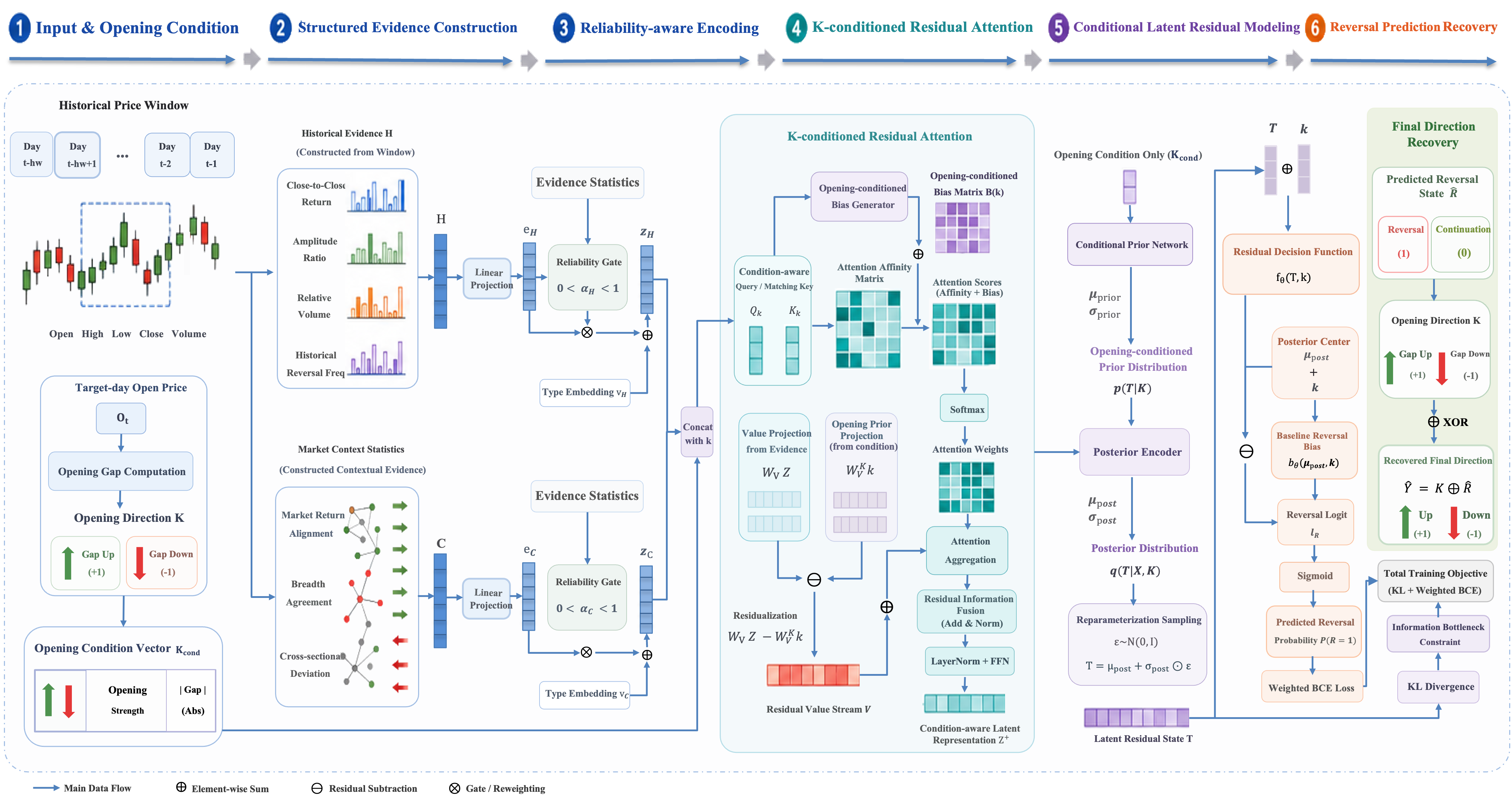}
  \caption{Overall architecture of \cope{}. The opening condition $K$ produces a reference center for the latent residual state $T$ via the conditional prior; historical state $H$ and context $C$ are reliability-gated and aggregated by $K$-conditioned residual attention into residual evidence $Z^+$; the reversal discrimination yields $\hat R$, and the adjusted-close direction prediction is recovered via $\hat Y=K\oplus\hat R$.}
  \label{fig:overall}
\end{figure}

\subsection{Conditional Reparameterization under the Opening State}

\cope{} treats the target-day opening state as an observed condition and learns whether the opening move is ultimately continued or corrected. Specifically, we define the conditional reversal label $R_{i,t}=Y_{i,t}\oplus K_{i,t}$: $R_{i,t}=0$ denotes continuation, and $R_{i,t}=1$ denotes reversal; the final direction is recovered via $Y_{i,t}=K_{i,t}\oplus R_{i,t}$. This transformation does not alter the label information; it only changes the parameterization of the conditional probability. For any evidence $X_{i,t}$ and opening condition $K_{i,t}$, we have
\begin{equation}
\begin{aligned}
p(Y_{i,t}\mid X_{i,t},K_{i,t})
&=\sum_{r\in\{0,1\}}p(Y_{i,t}\mid r,K_{i,t})p(r\mid X_{i,t},K_{i,t})\\
&=p(R_{i,t}=Y_{i,t}\oplus K_{i,t}\mid X_{i,t},K_{i,t}).
\end{aligned}
\label{eq:conditional_equivalence}
\end{equation}
Therefore, the core modeling object of \cope{} is
\begin{equation}
p_{\theta}(R_{i,t}\mid X_{i,t},K_{i,t}),
\end{equation}
where $K$ provides the opening-condition background and $X$ supplies the evidence needed to judge continuation or reversal. The subsequent model encodes residual information relative to the opening-conditioned prior and recovers the final direction after reversal decoding.
\subsection{Structured Evidence Decomposition}

\cope{} decomposes the evidence into $X_{i,t}=\{H_{i,t},C_{i,t}\}$, where $H$ captures the individual stock\'s historical tendency toward reversal and $C$ captures the cross-sectional environment\'s support or divergence from the opening move on the target day. The opening condition is separately encoded as
\begin{equation}
K^{\mathrm{cond}}_{i,t}=\left[\mathbf{1}(K_{i,t}=1),\mathbf{1}(K_{i,t}=0),\tilde g_{i,t},|\tilde g_{i,t}|\right],
\end{equation}
where $\tilde g$ is the standardized opening gap; the two binary dimensions indicate the opening direction, and the continuous dimensions capture the opening strength.

The individual historical state is defined as
\begin{equation}
H_{i,t}=\left[r^{\mathrm{cc}}_{i,t-1},q_{i,t-1},v_{i,t-1},\rho^{(hw)}_{i,t-1}\right],
\end{equation}
where $r^{\mathrm{cc}}$ is the previous day\'s close-to-close return, $q$ is the amplitude ratio, $v$ is the relative volume, and $\rho^{(hw)}$ is the raw reversal frequency within the historical window:
\begin{equation}
\rho^{(hw)}_{i,t-1}=\frac{1}{hw}\sum_{s=t-hw}^{t-1}R^{\mathrm{raw}}_{i,s}.
\end{equation}
The contextual support is denoted $C_{i,t}=\left[s^{\mathrm{ret}}_{i,t},s^{\mathrm{brd}}_{i,t},c^{\mathrm{crd}}_{i,t}\right]$, where $g_{i,t}$ is the raw opening gap, $s^{\mathrm{ret}}_{i,t}=g_{i,t}\bar r^{\mathrm{mkt}}_t$, $s^{\mathrm{brd}}_{i,t}=(2K_{i,t}-1)(2b^{\mathrm{mkt}}_t-1)$ measuring the agreement between the stock opening direction and market breadth direction, and $c^{\mathrm{crd}}_{i,t}=\left|g_{i,t}-\bar K^{\mathrm{mkt}}_t\right|$. All continuous evidence is standardized using training-set statistics, which are held fixed during validation and testing.
\subsection{Conditional Latent Residual Modeling}

The latent variable $T$ aggregates the posterior correction that $H$ and $C$ impose on the decision state under the opening condition. The conditional prediction is written as
\begin{equation}
p_{\theta}(R\mid X,K)=\int p_{\theta}(R\mid T,K)p(T\mid X,K)\,dT.
\end{equation}
Since the exact posterior is intractable, \cope{} approximates it with the variational posterior $q_{\phi}(T\mid X,K)$:
\begin{equation}
p_{\theta}(R\mid X,K)\approx\int p_{\theta}(R\mid T,K)q_{\phi}(T\mid X,K)\,dT.
\end{equation}

The opening-conditioned prior and approximate posterior are respectively defined as
\begin{equation}
p_{\psi}(T_{i,t}\mid K_{i,t})=\mathcal{N}\left(\mu_{\psi}(K^{\mathrm{cond}}_{i,t}),\operatorname{diag}(\sigma^2_{\psi}(K^{\mathrm{cond}}_{i,t}))\right),
\end{equation}
\begin{equation}
q_{\phi}(T_{i,t}\mid X_{i,t},K_{i,t})=\mathcal{N}\left(\mu_{\phi}(X_{i,t},K_{i,t}),\operatorname{diag}(\sigma^2_{\phi}(X_{i,t},K_{i,t}))\right).
\end{equation}
The former represents the latent decision center induced solely by the opening state, while the latter captures the sample-level correction contributed by the structured evidence. During training, reparameterized sampling is employed:
\begin{equation}
T_{i,t}=\mu_{\phi}(X_{i,t},K_{i,t})+\sigma_{\phi}(X_{i,t},K_{i,t})\odot\epsilon,
\qquad \epsilon\sim\mathcal{N}(0,I).
\end{equation}
\subsection{Reliability-Gated Evidence Tokens}

Before posterior encoding, the model maps each evidence group $g\in\{H,C\}$ to a token $e_g=E_g(x_g)$ and estimates its sample-level reliability. The group statistic vector is
\begin{equation}
s_g=\left[\eta_g,m_g,u_g,\log(1+d_g)\right],
\end{equation}
where $\eta_g$, $m_g$, $u_g$, and $d_g$ denote the nonzero ratio, mean absolute magnitude, standard deviation, and evidence dimensionality, respectively. The gating network outputs
\begin{equation}
\alpha_g=G_{\omega}(s_g),\qquad \alpha_g\in(0,1),
\end{equation}
and the gated evidence token is obtained as
\begin{equation}
z_g=\alpha_g e_g+\nu_g,
\end{equation}
where $\nu_g$ is the evidence-group type embedding. The tokens $z_H$ and $z_C$ then serve as inputs to the conditional attention aggregation.
\subsection{K-conditioned residual attention}

Given the token sequence $Z=[z_H,z_C]$ and the opening-condition embedding $k=e_K(K^{\mathrm{cond}}_{i,t})$, the posterior encoder models evidence-group interactions via K-conditioned residual attention and produces the parameters of $q_{\phi}(T\mid X,K)$. In each attention layer, the opening embedding is concatenated to every token to form condition-aware query and matching-key representations:
\begin{equation}
Q=W_Q[Z;k],\qquad M=W_M[Z;k].
\end{equation}
The value channel residualizes against the opening condition:
\begin{equation}
V=W_VZ-W_V^K\,k.
\label{eq:value_residual}
\end{equation}
Attention weights are further modulated by an opening-conditioned pairwise bias:
\begin{equation}
A=\operatorname{softmax}\left(\frac{QM^{\top}}{\sqrt{d_h}}+B(k)\right).
\end{equation}
The updated representation is
\begin{equation}
\widetilde Z=\operatorname{LN}\left(Z+W_O(AV)\right),\qquad
Z^+=\operatorname{LN}\left(\widetilde Z+\operatorname{FFN}(\widetilde Z)\right).
\end{equation}
Here $W_V^K$ is the opening-condition-specific weight in the value channel, with $W_V^Kk$ broadcast across the token dimension. The conditioned query/key determines how evidence groups retrieve one another, and value residualization ensures that the transmitted information reflects the residual explanatory content relative to the opening condition. Finally, $Z^+$ is aggregated to obtain $\mu_{\phi}(X_{i,t},K_{i,t})$ and $\sigma_{\phi}(X_{i,t},K_{i,t})$.
\subsection{Training Objective and Inference}

During training, the model first samples the latent residual state $T$ from the variational posterior and then performs reversal discrimination jointly with the opening-condition embedding $k$. The reversal decoder outputs the logit
\begin{equation}
\ell_R=f_{\theta}(T,k)-b_{\theta}(\mu_{\phi},k),\qquad
\hat p_R=\sigma(\ell_R).
\end{equation}
Here $b_{\theta}(\mu_{\phi},k)$ models the baseline reversal tendency induced by the opening condition and the posterior center, enabling the main decoder to focus on sample-level residual evidence. At test time, the posterior mean is used for deterministic inference:
\begin{equation}
\ell_R^{\mathrm{test}}=f_{\theta}(\mu_{\phi},k)-b_{\theta}(\mu_{\phi},k).
\end{equation}
The final decision is
\begin{equation}
\hat R=\mathbf{1}(\hat p_R\ge 0.5),\qquad \hat Y=K\oplus\hat R.
\end{equation}

The training objective adopts a discriminative information bottleneck referenced to the opening condition:
\begin{equation}
\mathcal{L}=\widetilde{\mathcal{L}}_{\mathrm{KL}}+\beta_e\mathcal{L}_{\mathrm{BCE}}.
\end{equation}
The KL term constrains the posterior residual state to stay close to the opening-conditioned prior:
\begin{equation}
\mathcal{L}_{\mathrm{KL}}=\frac{1}{N}\sum_{i=1}^{N}\operatorname{KL}\left(q_{\phi}(T_{i,t}\mid X_{i,t},K_{i,t})\Vert p_{\psi}(T_{i,t}\mid K_{i,t})\right).
\end{equation}
To prevent early KL collapse, a free-bits correction is applied:
\begin{equation}
\widetilde{\operatorname{KL}}_{i,t}=d_T\cdot\max\left(\frac{1}{d_T}\operatorname{KL}_{i,t},\lambda_{\mathrm{fb}}\right).
\end{equation}
The reversal BCE uses class weights to mitigate the scarcity of reversal samples:
\begin{equation}
\mathcal{L}_{\mathrm{BCE}}=-\frac{1}{N}\sum_{i=1}^{N}w_{i,t}\left[R_{i,t}\log\hat p_{R,i,t}+(1-R_{i,t})\log(1-\hat p_{R,i,t})\right],
\end{equation}
where $w_{i,t}=N_{\mathrm{cont}}/N_{\mathrm{rev}}$ when $R_{i,t}=1$ and $w_{i,t}=1$ otherwise; the warmup coefficient is $\beta_e=\beta\min\left(1,\frac{e}{E_{\mathrm{warm}}}\right)$.

\section{Experiments}
\label{sec:experiments}

\subsection{Experimental Setup}

\begin{table*}[t]
\centering
\scriptsize
\setlength{\tabcolsep}{3.4pt}
\renewcommand{\arraystretch}{1.1}
\resizebox{0.95\linewidth}{!}{
\begin{tabular}{@{}ll|>{\centering\arraybackslash}p{1cm}|cccc@{}}
\toprule
Category & Method & Metric & ACL18 & CMIN-US & CMIN-CN & KDD17 \\
\midrule
\multirow{14}{*}{Multimodal} & \multirow{2}{*}{StockNet} & \metriccell{ACC} & 58.23 $\pm$ 0.0030 & 52.46 $\pm$ 0.0041 & 54.53 $\pm$ 0.0062 & 51.93 $\pm$ 0.0001 \\
& & \metriccell{MCC} & 0.081 $\pm$ 0.0071 & 0.022 $\pm$ 0.0025 & 0.045 $\pm$ 0.0043 & 0.034 $\pm$ 0.0050 \\
 & \multirow{2}{*}{CMIN} & \metriccell{ACC} & 62.69 $\pm$ 0.0029 & 53.43 $\pm$ 0.0085 & 55.28 $\pm$ 0.0094 & -- \\
& & \metriccell{MCC} & 0.209 $\pm$ 0.0016 & 0.046 $\pm$ 0.0055 & 0.111 $\pm$ 0.0990 & -- \\
 & \multirow{2}{*}{CausalStock} & \metriccell{ACC} & 63.42 $\pm$ 0.0039 & 54.64 $\pm$ 0.0083 & 56.19 $\pm$ 0.0084 & \textcolor{blue}{56.09 $\pm$ 0.0069} \\
& & \metriccell{MCC} & 0.217 $\pm$ 0.0017 & 0.048 $\pm$ 0.0057 & 0.142 $\pm$ 0.0813 & 0.124 $\pm$ 0.0189 \\
 & \multirow{2}{*}{MagicNet} & \metriccell{ACC} & 63.74 $\pm$ 0.0035 & 55.12 $\pm$ 0.0070 & 56.45 $\pm$ 0.0080 & -- \\
& & \metriccell{MCC} & \textcolor{blue}{0.231 $\pm$ 0.0030} & 0.086 $\pm$ 0.0060 & 0.134 $\pm$ 0.0080 & -- \\
 & \multirow{2}{*}{HAN} & \metriccell{ACC} & 57.64 $\pm$ 0.0040 & 53.72 $\pm$ 0.0020 & 53.59 $\pm$ 0.0037 & 54.58 $\pm$ 0.0060 \\
& & \metriccell{MCC} & 0.052 $\pm$ 0.0050 & 0.100 $\pm$ 0.0015 & 0.016 $\pm$ 0.0026 & 0.055 $\pm$ 0.0120 \\
 & \multirow{2}{*}{PEN} & \metriccell{ACC} & 59.89 $\pm$ 0.0090 & 53.20 $\pm$ 0.0051 & 54.83 $\pm$ 0.0086 & 55.69 $\pm$ 0.0070 \\
& & \metriccell{MCC} & 0.156 $\pm$ 0.0018 & 0.027 $\pm$ 0.0023 & 0.086 $\pm$ 0.0065 & \textcolor{blue}{0.137 $\pm$ 0.0150} \\
 & \multirow{2}{*}{LLMFactor} & \metriccell{ACC} & \textcolor{blue}{65.81 $\pm$ 0.0045} & \textcolor{blue}{61.71 $\pm$ 0.0065} & \textcolor{blue}{60.59 $\pm$ 0.0070} & -- \\
& & \metriccell{MCC} & 0.228 $\pm$ 0.0040 & \textcolor{blue}{0.228 $\pm$ 0.0060} & \textcolor{blue}{0.245 $\pm$ 0.0080} & -- \\
\midrule
\multirow{8}{*}{Unimodal} & \multirow{2}{*}{ALSTM} & \metriccell{ACC} & 51.81 $\pm$ 0.0050 & 51.64 $\pm$ 0.0060 & 53.35 $\pm$ 0.0045 & 51.66 $\pm$ 0.0041 \\
& & \metriccell{MCC} & 0.032 $\pm$ 0.0120 & 0.006 $\pm$ 0.0140 & 0.023 $\pm$ 0.0080 & 0.032 $\pm$ 0.0119 \\
 & \multirow{2}{*}{LSTM} & \metriccell{ACC} & 51.31 $\pm$ 0.0060 & 54.15 $\pm$ 0.0070 & 52.07 $\pm$ 0.0060 & 51.18 $\pm$ 0.0066 \\
& & \metriccell{MCC} & 0.021 $\pm$ 0.0110 & 0.042 $\pm$ 0.0160 & 0.028 $\pm$ 0.0130 & 0.019 $\pm$ 0.0110 \\
 & \multirow{2}{*}{Adv-ALSTM} & \metriccell{ACC} & 52.75 $\pm$ 0.0055 & 51.73 $\pm$ 0.0070 & 53.49 $\pm$ 0.0050 & 51.69 $\pm$ 0.0058 \\
& & \metriccell{MCC} & 0.052 $\pm$ 0.0140 & 0.012 $\pm$ 0.0200 & 0.025 $\pm$ 0.0140 & 0.033 $\pm$ 0.0137 \\
 & \multirow{2}{*}{DTML} & \metriccell{ACC} & 57.44 $\pm$ 0.0075 & 52.06 $\pm$ 0.0103 & 54.42 $\pm$ 0.0121 & 53.53 $\pm$ 0.0075 \\
& & \metriccell{MCC} & 0.191 $\pm$ 0.0195 & 0.031 $\pm$ 0.0230 & 0.083 $\pm$ 0.0214 & 0.073 $\pm$ 0.0195 \\
\midrule
\multirow{2}{*}{Ours} & \multirow{2}{*}{\cope} & \metriccell{ACC} & \textbf{75.11 $\pm$ 0.0142} & \textbf{75.70 $\pm$ 0.0094} & \textbf{71.43 $\pm$ 0.0042} & \textbf{84.30 $\pm$ 0.0070} \\
& & \metriccell{MCC} & \textbf{0.500 $\pm$ 0.0286} & \textbf{0.511 $\pm$ 0.0193} & \textbf{0.427 $\pm$ 0.0082} & \textbf{0.686 $\pm$ 0.0144} \\
\bottomrule
\end{tabular}}
\caption{Main results on four stock prediction datasets. The table reports Accuracy (ACC) and Matthews Correlation Coefficient (MCC).}
\label{tab:main}
\end{table*}

We evaluate our method on four datasets spanning different markets: ACL18\footnote{\url{https://github.com/yumoxu/stocknet-dataset}}, CMIN-US\footnote{\label{fn:cmin_dataset}\url{https://github.com/BigRoddy/CMIN-Dataset}}, CMIN-CN\footref{fn:cmin_dataset}, and KDD17~\cite{zhang2017stock}\footnote{\url{https://github.com/fulifeng/Adv-ALSTM/tree/master/data/kdd17}}. All dataset links point to public data sources. Main comparison results are reported as the mean and standard deviation over 5 random seeds under a unified data split, label definition, and sample-filtering protocol. All experiments were conducted on NVIDIA A100 GPUs. Additional experimental and implementation details are provided in the appendix.

Baselines include unimodal price models such as ALSTM~\cite{qin2017dual}, Adv-ALSTM~\cite{ijcai2019p810}, and DTML~\cite{yoo2021accurate}, as well as multimodal or relation-modeling methods such as StockNet~\cite{xu2018stock}, CMIN~\cite{luo2023causality}, CausalStock~\cite{li2024causalstock}, MagicNet~\cite{luo2025magicnet}, HAN~\cite{hu2018listening}, PEN~\cite{li2023pen}, and LLMFactor~\cite{wang2024llmfactor}. We report ACC and MCC as the two primary metrics under identical evaluation rules. Shortcut analysis reports two metrics:
\begin{equation}
\mathrm{SDR}=\frac{1}{N}\sum_{i,t}\mathbf{1}(\hat Y_{i,t}=K_{i,t}),
\end{equation}
i.e., the proportion of predictions that agree with the opening direction, and
\begin{equation}
\rho=\max\left(0,\frac{\mathrm{SDR}-0.5}{\mathrm{ACC}-0.5+\varepsilon}\right).
\end{equation}
A higher $\rho$ indicates that the model\'s above-chance performance is more strongly associated with opening-direction copying.

\subsection{Main Results}

Table~\ref{tab:main} presents the main results on four datasets. \cope{} achieves the best ACC and MCC across all datasets, demonstrating stable cross-market predictive capability. Compared with unimodal price models, \cope{} explicitly models the continuation/reversal relationship after the target-day open, thereby leveraging the new information available at the opening time more effectively. Compared with multimodal methods that rely on news or cross-stock relations, \cope{} does not introduce additional modalities but instead reorganizes existing price information through conditional modeling. The concurrent improvement in MCC further indicates that the model\'s advantage stems not from class bias but from more reliable directional discrimination. The main comparison results validate the significant effectiveness of using opening information as a conditional discrimination anchor.

\subsection{Ablation Study}

\begin{table}[t]
\centering
\scriptsize
\setlength{\tabcolsep}{2.2pt}
\renewcommand{\arraystretch}{1.08}
\resizebox{0.9\linewidth}{!}{%
\begin{tabular}{l|cc|cc|cc|cc}
\toprule
\multirow{2}{*}{Model Variant} & \multicolumn{2}{c|}{ACL18} & \multicolumn{2}{c|}{CMIN-US} & \multicolumn{2}{c|}{CMIN-CN} & \multicolumn{2}{c}{KDD17} \\
\cmidrule(lr){2-3}\cmidrule(lr){4-5}\cmidrule(lr){6-7}\cmidrule(lr){8-9}
& ACC & $\rho$ & ACC & $\rho$ & ACC & $\rho$ & ACC & $\rho$ \\
\midrule
\cope & \textbf{75.11} & \textbf{0.498} & \textbf{75.70} & \textbf{0.428} & \textbf{71.43} & \textbf{0.083} & \textbf{84.30} & \textbf{0.584} \\
w/o conditional prior & 74.49 & 0.594 & 75.61 & 0.575 & 70.76 & 0.315 & 83.90 & 0.714 \\
w/o reliability gating & 73.45 & 0.517 & 76.08 & 0.610 & 70.91 & 0.301 & 83.33 & 0.648 \\
w/o K-conditioned residual attention & 74.52 & 0.558 & 75.03 & 0.536 & 70.31 & 0.246 & 83.18 & 0.666 \\
w/o contextual support & 56.86 & 0.000 & 62.07 & 0.197 & 55.92 & 0.000 & 67.81 & 0.626 \\
direct opening input variant & 73.05 & 2.057 & 54.04 & 1.802 & 61.58 & 3.700 & 50.86 & 1.316 \\
\bottomrule
\end{tabular}}
\caption{Ablation results. The table reports ACC and shortcut dependence ratio $\rho$ for each model variant on four datasets.}
\label{tab:ablation}
\end{table}

Table~\ref{tab:ablation} summarizes the contribution of each component. Removing contextual support leads to the largest ACC degradation, showing that market context provides essential evidence for distinguishing continuation from reversal after the open. Removing the conditional prior, reliability gating, or K-conditioned residual attention has a milder effect on ACC, but consistently increases $\rho$ on most datasets; these modules therefore mainly help prevent the model from relying on direct opening-direction copying. The direct opening input variant further supports this interpretation: although it can obtain moderate ACC in some cases, its shortcut dependence is much stronger. Overall, the ablation results indicate that contextual support supplies discriminative information, while the prior, gating, and conditioned attention improve robustness against opening-price shortcuts.

\subsection{Shortcut-Sensitivity Analysis}

\begin{table*}[t]
\centering
\scriptsize
\setlength{\tabcolsep}{2.2pt}
\renewcommand{\arraystretch}{1.06}
\resizebox{0.9\linewidth}{!}{%
\begin{tabular}{l|ccc|ccc|ccc|ccc}
\toprule
\multirow{2}{*}{Method} & \multicolumn{3}{c|}{ACL18} & \multicolumn{3}{c|}{CMIN-US} & \multicolumn{3}{c|}{CMIN-CN} & \multicolumn{3}{c}{KDD17} \\
\cmidrule(lr){2-4}\cmidrule(lr){5-7}\cmidrule(lr){8-10}\cmidrule(lr){11-13}
& ACC & SDR & $\rho$ & ACC & SDR & $\rho$ & ACC & SDR & $\rho$ & ACC & SDR & $\rho$ \\
\midrule
RNN+Opening & 73.24 & 0.890 & 1.678 & 69.15 & 0.832 & 1.733 & 60.21 & 0.770 & 2.644 & 76.58 & 0.924 & 1.596 \\
LSTM+Opening & 55.04 & 0.543 & 0.855 & 54.04 & 0.573 & 1.802 & 49.62 & 0.477 & 0.000 & 49.14 & 0.489 & 0.000 \\
Transformer+Opening & 73.05 & 0.974 & 2.057 & 54.04 & 0.573 & 1.802 & 61.58 & 0.929 & 3.700 & 50.86 & 0.511 & 1.316 \\
\cope & \textbf{75.11} & \textbf{0.625} & \textbf{0.498} & \textbf{75.70} & \textbf{0.610} & \textbf{0.428} & \textbf{71.43} & \textbf{0.518} & \textbf{0.083} & \textbf{84.30} & \textbf{0.700} & \textbf{0.584} \\
\bottomrule
\end{tabular}}
\caption{Shortcut-sensitivity analysis. The table compares direct-opening-input models and \cope{} on four datasets in terms of ACC, SDR, and $\rho$.}
\label{tab:shortcut}
\end{table*}

Table~\ref{tab:shortcut} further compares the shortcut-sensitivity metrics between direct-opening-input models and \cope{}. Consistent with the pre-experiment, conventional models achieve relatively high ACC on some datasets when the opening-price shortcut is not accounted for, but this comes with elevated SDR and $\rho$, indicating that much of their performance derives from copying the target-day opening direction rather than effective learning. In contrast, \cope{} maintains the highest ACC across all four datasets while keeping $\rho$ at substantially lower levels, demonstrating its advantage in balancing opening-information utilization and shortcut-copying suppression. LSTM+Opening remains near chance level on several datasets, which also shows that a low SDR by itself does not indicate model robustness---its true cause is the inability to handle the shortcut problem and to learn an effective decision boundary.

\subsection{Attention Visualization}

\begin{figure}[t]
  \centering
  \includegraphics[width=\linewidth]{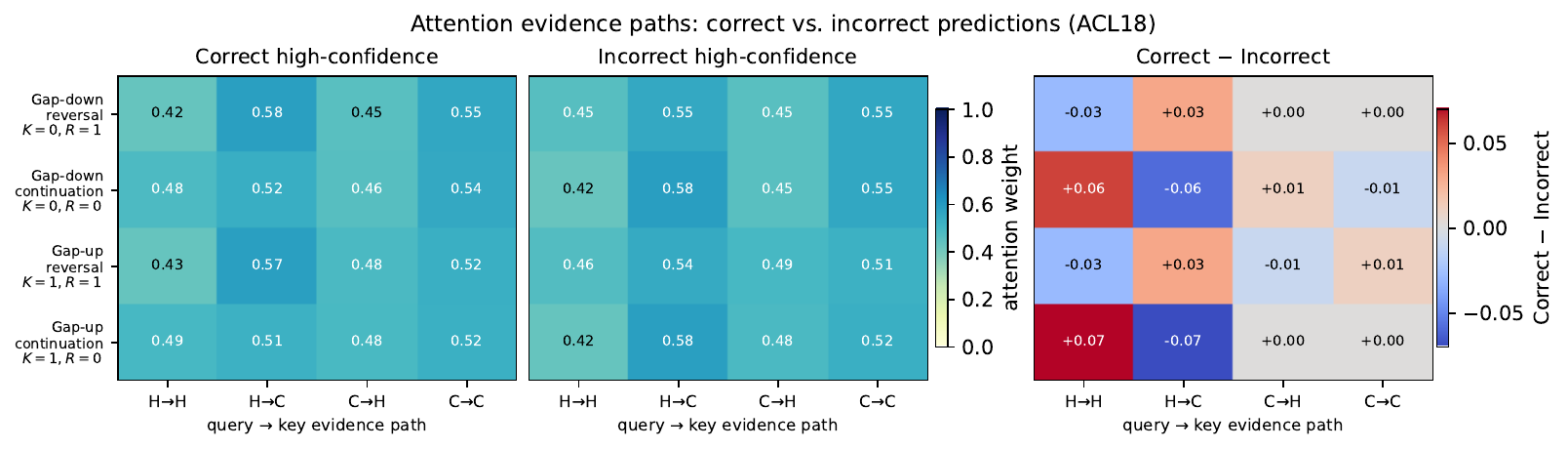}
  \caption{Correctness-contrast visualization of K-conditioned residual attention. Left and center panels show average attention weights for high-confidence correct and incorrect predictions, respectively; the right panel shows their difference.}
  \label{fig:attention}
\end{figure}

Figure~\ref{fig:attention} shows the average attention patterns for high-confidence correct and incorrect predictions in the ACL18 test set. Samples are partitioned by $(K,R)$, and four attention evidence paths between history evidence and market context are computed. Correct reversal predictions emphasize the $H\rightarrow C$ cross-evidence path, while correct continuation predictions emphasize the $H\rightarrow H$ historical self-attention path; in contrast, attention distributions for incorrect predictions are more dispersed. This result indicates that K-conditioned residual attention can dynamically select evidence sources depending on whether the post-open state is continuation or reversal, enabling \cope{} to not only utilize opening information itself but also extract residual information relative to the opening prior, thereby improving prediction reliability.

\subsection{Simulated Investment and Backtesting}

\noindent
\raisebox{0pt}[0pt][0pt]{\makebox[\linewidth][r]{%
\begin{minipage}[t]{0.52\linewidth}
\vspace{0pt}
\centering
\scriptsize
\setlength{\tabcolsep}{2.0pt}
\renewcommand{\arraystretch}{1.06}
\resizebox{\linewidth}{!}{%
\begin{tabular}{llrrrr}
\toprule
Method & Metric & ACL18 & CMIN-US & CMIN-CN & KDD17 \\
\midrule
\multirow{3}{*}{History-only} & Ann. Return & \textcolor{blue}{-5.31} & \textcolor{blue}{-24.95} & \textcolor{blue}{98.13} & \textcolor{blue}{-17.05} \\
& Sharpe Ratio & \textcolor{blue}{-0.100} & \textcolor{blue}{-1.879} & \textcolor{blue}{2.484} & \textcolor{blue}{-1.241} \\
& Max Drawdown & \textcolor{blue}{21.69} & \textcolor{blue}{38.01} & \textcolor{blue}{16.36} & \textcolor{blue}{31.92} \\
\midrule
\multirow{3}{*}{\cope{}} & Ann. Return & \textbf{24.46} & \textbf{41.16} & \textbf{147.30} & \textbf{49.54} \\
& Sharpe Ratio & \textbf{1.111} & \textbf{2.601} & \textbf{3.959} & \textbf{3.141} \\
& Max Drawdown & \textbf{14.90} & \textbf{7.28} & \textbf{7.93} & \textbf{6.29} \\
\bottomrule
\end{tabular}}
\refstepcounter{table}\label{tab:backtest}
\par\vspace{0.3em}
{\small Table~\thetable: Low-open reversal long-only backtest results under the unified protocol. The table reports annualized return, Sharpe ratio, and maximum drawdown.}
\end{minipage}}}
\parshape=16
0pt 0.46\linewidth
0pt 0.46\linewidth
0pt 0.46\linewidth
0pt 0.46\linewidth
0pt 0.46\linewidth
0pt 0.46\linewidth
0pt 0.46\linewidth
0pt 0.46\linewidth
0pt 0.46\linewidth
0pt 0.46\linewidth
0pt 0.46\linewidth
0pt 0.46\linewidth
0pt 0.46\linewidth
0pt 0.46\linewidth
0pt \linewidth
0pt \linewidth
{\small Table~\ref{tab:backtest} presents low-open reversal backtest results under the unified protocol. The strategy retains only low-open stocks after the open, ranks them by reversal confidence, and holds to the close; the full protocol is in Appendix~\ref{app:backtest_protocol}. History-only is a baseline without target-day opening state. \cope{} outperforms History-only in annualized return, Sharpe ratio, and maximum drawdown, indicating that its reversal confidence provides more effective ranking signals in opening-decision scenarios.}\par
\vspace{0.8em}

\section{Conclusion}
\label{sec:conclusion}

This paper formalizes the opening-price shortcut as a measurable risk in adjusted-close direction prediction and proposes \cope{}. \cope{} first models the opening state as an observed condition and reparameterizes direction prediction into conditional continuation/reversal discrimination; it then encodes only residual evidence relative to the opening-conditioned prior, suppressing opening-direction copying through reliability gating and K-conditioned residual attention; the conditional information bottleneck further constrains the residual state from deviating excessively from the opening prior. Experiments on four datasets confirm that \cope{} consistently improves predictive performance, significantly suppresses the opening-direction copying shortcut, and shows strong application value in opening-decision backtests. This work provides a conditional modeling approach for identifying and mitigating the opening-price shortcut.

\clearpage
\appendix

\section{Theoretical Details}
\label{app:theory}

\subsection{Label Equivalence of Conditional Reversal Reparameterization}
For any $Y,K\in\{0,1\}$, define $R=Y\oplus K$. Since XOR is self-inverse,
\begin{equation}
K\oplus R=K\oplus(Y\oplus K)=Y\oplus(K\oplus K)=Y.
\end{equation}
Hence, predicting $R$ and recovering $Y$ via $K\oplus R$ is label-equivalent to directly predicting $Y$.

\subsection{Conditional Distribution Equivalence}
Since $R=Y\oplus K$ is a deterministic transformation,
\begin{equation}
p(Y=y\mid X,K)=p(R=y\oplus K\mid X,K).
\end{equation}
Therefore, optimizing the reversal likelihood does not alter the original closing-direction prediction target; it only changes the parameterization of the conditional probability.

\subsection{Shortcut Risk of Directly Inputting the Opening Direction}
If the empirical distribution satisfies $P(Y=K)>0.5$, the classifier $h_K(X,K)=K$ achieves accuracy
\begin{equation}
\operatorname{Acc}(h_K)=P(Y=K).
\end{equation}
When this value is high, an empirical risk minimizer can rely on a low-complexity shortcut to achieve strong surface performance, thereby reducing its use of evidence that genuinely discriminates continuation from reversal.

\subsection{Validity of SDR and $\rho$}
For any model, $\mathrm{SDR}=P(\hat Y=K)$ directly measures the agreement between the predicted direction and the opening direction. When $\mathrm{ACC}>0.5$,
\begin{equation}
\rho_{\mathrm{signed}}=\frac{\mathrm{SDR}-0.5}{\mathrm{ACC}-0.5+\varepsilon}
\end{equation}
compares above-chance opening agreement with above-chance predictive accuracy. The main text reports $\rho=\max(0,\rho_{\mathrm{signed}})$ and requires that it be interpreted alongside ACC and MCC.

\section{Baseline Details}
\label{app:baselines}
Baselines in the main comparison fall into two categories. The first category comprises unimodal models that use only historical price sequences, including ALSTM, Adv-ALSTM, and DTML, which enhance price-sequence representations through attention recurrent networks, adversarial training, or trend modeling, respectively. The second category includes multimodal or relation-modeling methods that incorporate news, events, cross-stock relations, or language model factors, including StockNet, CMIN, CausalStock, MagicNet, HAN, PEN, and LLMFactor. To ensure fair comparison, we re-report ACC and MCC for all baselines under a unified data split, label definition, and sample-filtering protocol.

The shortcut-diagnosis models RNN+Opening, LSTM+Opening, and Transformer+Opening are direct-opening variants that add target-day opening-direction or opening-state features to the conventional historical price input. They are used to diagnose whether directly concatenating opening information induces the model to copy the opening direction, rather than serving as ablation components of the proposed framework.

\section{Metric Definitions}
\label{app:metrics}
ACC and MCC are computed from true positives, true negatives, false positives, and false negatives:
\begin{equation}
\mathrm{ACC}=\frac{tp+tn}{tp+tn+fp+fn},
\end{equation}
\begin{equation}
\mathrm{MCC}=\frac{tp\cdot tn-fp\cdot fn}{\sqrt{(tp+fp)(tp+fn)(tn+fp)(tn+fn)}}.
\end{equation}
ACC measures overall classification accuracy, while MCC accounts for all four confusion-matrix entries and is more suitable for evaluating binary classification when class distributions are imbalanced. SDR, defined in the main text, measures the proportion of predictions that agree with the target-day opening direction, and $\rho$ quantifies the strength of this agreement relative to above-chance accuracy gains. Thus, ACC/MCC evaluate final direction-prediction capability, while SDR/$\rho$ diagnose whether predictions over-rely on opening-direction copying.

\section{Dataset Details}
\label{app:data}
This paper uses four public stock prediction datasets---ACL18, CMIN-US, CMIN-CN, and KDD17---covering different market environments including the US and China. All datasets are split chronologically into training, validation, and test sets, with a unified label definition: the target is the direction of the adjusted close relative to the previous day\'s adjusted close, and the target-day opening direction is determined from the raw opening price relative to the previous day\'s raw closing price. If the previous day\'s closing price is missing, or the absolute change in the target-day adjusted close relative to the previous day falls below the minimum threshold of $0.003$, the stock--day sample is excluded from the supervision set. All standardization statistics are estimated solely from the training set and held fixed during validation and testing to prevent test-set information leakage.

\section{Implementation Details}
\label{app:implementation}
The model standardizes continuous features using training-set statistics and applies the same transformation during validation and testing. Training jointly optimizes binary cross-entropy and the conditional information bottleneck term; reversal class weights mitigate the continuation/reversal distribution imbalance, while free bits and supervised warmup prevent the KL constraint from becoming too strong in early training. All main comparison results report the mean and standard deviation over 5 random seeds.

Checkpoint selection uses the composite validation score:
\begin{equation}
\mathrm{Score}=\mathrm{ACC}+0.20\,\mathrm{MCC}+0.20\,A_{\mathrm{bal}}-0.05\,|\rho_{\mathrm{signed}}|,
\end{equation}
where $A_{\mathrm{bal}}=\min(A_{\mathrm{cont}},A_{\mathrm{rev}})$. This score serves solely as an implementation detail for model selection and is not a method objective of the main text, nor does it participate in test-set evaluation.

\section{Investment Simulation Protocol}
\label{app:backtest_protocol}

The purpose of the investment simulation is not to construct a complete trading system, but to test whether the opening-conditioned reversal signal proposed in this paper exhibits economic consistency under a unified, leak-free, and executable ranking protocol. Since opening information can be used both for adjusted-close direction prediction and for post-open trading decisions, mixing different return conventions would significantly alter backtest conclusions. Therefore, the main text reports only the protocol consistent with the low-open reversal objective and excludes other diagnostic backtests from the main results.

Specifically, on each test trading day, all methods first observe the target-day opening state, historical state, and market context. The backtest then retains only low-open stocks as the candidate pool, because this scenario directly corresponds to the trading question of ``will a reversal occur after the open?'' For each candidate stock, \cope{} uses the reversal probability as the ranking score, while History-only uses a historical-feature model without the target-day opening condition as the comparison score. Each day, the top 20\% of stocks by score are selected, held with equal weight, and liquidated at the close. Daily returns are computed using open-to-close returns minus a uniform transaction cost; annualized return, Sharpe ratio, and maximum drawdown are all derived from the test-period daily return series.

To ensure fair comparison, \cope{} and History-only use the same stock pool, same test dates, same candidate set, same transaction cost, same top-20\% selection ratio, and same equal-weight portfolio rule. All model selection and standardization parameters are determined solely from training/validation data; the test set is used only for final evaluation. The target-day closing price does not participate in signal generation and is used only to compute realized returns after the backtest concludes. Therefore, this protocol does not leak unobservable closing information into the opening-time decision process.

Earlier diagnostic backtest results were unstable, primarily because different experiments effectively evaluated different economic questions. For example, if adjusted-close direction probabilities are used to rank stocks but open-to-close returns are used to compute portfolio returns, the model score corresponds to direction relative to the previous day\'s adjusted close, whereas the portfolio return reflects post-open price change---the two objectives are misaligned. This paper ultimately adopts the low-open reversal protocol because it aligns the model output, the candidate stock set, and the return computation around the same opening-conditioned reversal event. Table~\ref{tab:backtest} in the main text therefore reports only the results under this frozen protocol. It should be emphasized that this backtest still does not model real-market slippage, liquidity constraints, market-impact costs, or capacity limits, and should therefore be understood as a test of economic consistency for the predictive signal rather than a deployable investment strategy.

\section{Societal Impact}
\label{app:impact}
Financial prediction models can be used for research, risk analysis, and investment assistance, but they may also be misused as overconfident trading signals or misleading investment promises. The backtest results reported in this paper do not eliminate market risk, liquidity risk, transaction costs, capacity limits, or execution deviations. Our method should be understood as a predictive modeling framework, not investment advice.

\section{Limitations}
\label{app:limitations}
As the first work to directly analyze the impact of the opening state on the closing direction, this paper focuses on the price-data microstructure rather than multimodal features. This is necessary for fully understanding opening information; after the price-internal opening characteristics are more comprehensively captured, future work can incorporate overnight news, announcements, social media, or macroeconomic events as additional modalities. Constrained by the overnight window size, such extensions require minute- or hourly-level news timestamps and strict controllability of information availability, which are beyond the scope of this paper.

Furthermore, our evaluation is based on daily-frequency direction prediction and historical backtests. Real trading is affected by slippage, market-impact costs, liquidity constraints, trading regulations, portfolio capacity, and risk-management rules; hence, backtest performance should not be directly interpreted as executable returns. Future work may evaluate the trading robustness of this framework under more realistic matching and cost models.

\clearpage
\bibliographystyle{plain}
\bibliography{custom}

\end{document}